9/24/20

# Using math in physics: Overview

*Edward F. Redish,*
University of Maryland - emeritus, College Park, MD

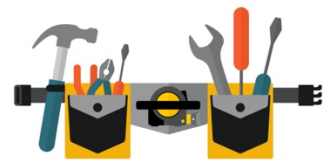

The key difference between math as math and math in science is that in science we blend our physical knowledge with our knowledge of math. This blending changes the way we put meaning to math and even to the way we interpret mathematical equations. Learning to think about physics with math instead of just calculating involves a number of general scientific thinking skills that are often taken for granted (and rarely taught) in physics classes. In this paper, I give an overview of my analysis of these additional skills. I propose specific tools for helping students develop these skills in subsequent papers.

Many of the ideas and methods I'm discussing here were developed in the context of studying introductory physics with life science students — first, in algebra-based physics and then in NEXUS/Physics, an introductory physics course designed specifically for life science majors. Students in these classes often struggle with the idea that symbolic quantities in science represent physical measurements rather than numbers and that equations represent relationships rather than ways to calculate.

## Math in science is different from math in math.

In science, symbols stand for *a blend* — a mental combination of physical knowledge with knowledge of how a mathematical element such as a variable or constant behaves. This changes the way we think about and use equations. For example, when we define the electric field as $E = F/q$ we have in mind that $F$ is not just an arbitrary variable but the specific electric force felt by the test charge $q$, a conceptual blend of physics and math. In math, we would include the q-dependence explicitly in our label. In physics, we do not. Rather, we expect the viewer to interpret the symbol as something physical and therefore to realize that when $q$ changes, so does $F$. As a result, when $q$ changes, $E$ does not change, surprising students.

- Math in math classes tends to be about numbers. Math in science is not. Math in science blends physics conceptual knowledge with mathematical symbols.

Math in science is about relations among physical quantities that are transformed into numbers by *measurement*. As a result, quantities in science tend to have dimensions and units. These have to be treated differently from ordinary numbers.

Unlike ordinary numbers, different kinds of quantities can't be equated. Students wonder why equations like $x = t$ (and 3 cm = 3 seconds) are forbidden but the equation 2.54 cm = 1 inch is allowed. I'll discuss this in detail in the paper in this series on Dimensional Analysis.[1].

Students don't usually learn to do this blending in math classes, and most students in introductory physics have no experience with it.[2] This blending has a lot of structure and results in differences in the ways we use symbols in math and science.

- Math in math classes tends to use a small number of symbols in constrained ways. Math in physics uses *lots* of symbols in different ways - and the same symbol may have different meanings.

In a typical algebra or calculus book, you will find very few equations with more than one or two symbols and they tend to follow a predictable convention — x, y, z, and t will be variables; f, g, and h will be functions; a, b, and c will be constants. In a typical physics book, you will rarely find an equation with fewer than 3 symbols and you will often find ones with 6 or more. And they won't follow the math conventions. This makes the equations we use in physics look unfamiliar to students and raises their level of discomfort.

- The symbols in science classes often carry meaning that changes the way we interpret the quantity.

In pure math it doesn't matter what we call something; in science it does. In science, we choose a symbol for a variable or constant to give us a hint as to what kind of quantity we are talking about. We use *m* for a mass and *t* for a time — never the other way round. Even more confusing is the fact that we use the *same* symbol to mean different things. In my class, the symbol $Q$ can stand for heat, electric charge, or volumetric flow. $T$ can stand for a tension, a temperature, or the period of an oscillation.

You might say, "Well, sure. But the interpretation depends on the context. Then it's obvious what you mean." True. But looking for the context means that you are already **blending** your knowledge of what the symbols mean physically with your mathematical knowledge. Equations in physics not only represent quantitative knowledge of the physical world. Through the blend, they codify both physical conceptual knowledge and functional dependence. I discuss these issues in more detail in the papers in this series on Anchor Equations[3] and Functional Dependence.[4]

- In introductory math, symbols tend to stand for either variables or constants. In science, we have lots of different kinds of symbols and they may shift from constant to variable, depending on what we want to do.

Math in math seems so clean. A variable is a variable. A constant is a constant. In physics, our constants can be universal constants, parameters, initial conditions; and we might choose to differentiate with respect to them.[5]

- In introductory math, equations are almost always about solving and calculating. In physics it's often about *explaining*.

Every semester, I have one or two students whose comments go something like, "He doesn't explain enough. He spends too much time doing algebra on the board." These students haven't picked up on something I try to make explicit (but obviously don't succeed in for all students): A derivation of an equation *is* an explanation if you see the equations as carrying conceptual meaning. This is something that is not obvious and is not easy to learn, especially when their other science classes are not using math in this way.

These "student difficulties" are different from the usual student ones we might worry about. They're not just lack of knowledge, like forgetting how to divide by a fraction. These are more complex and "meta" — another level, overlaying everything the students do. A lot of the literature in physics education is about helping students better understand physics concepts. A focus on concepts is sometimes looked at as getting away from the math. But how can we think about student thinking when the concept to be learned is fundamentally mathematical as well as physical?

## Am I being unfair to math?

I know some mathematical physicists (and perhaps even some mathematicians) are going to complain that I've made math into a straw man. Many of the things I say that math doesn't do, it of course does — in more advanced classes. In introductory physics, there's some interesting "hidden" advanced math.[6] We use some very sophisticated mathematical structures in an introductory class because we expect our students to interpret them not using fancy math but through the blend with common physical knowledge.

For example, the fact that the dimensions of an equation have to match can be seen as a group theoretical statement: since the measurement scales are arbitrary, the equations must transform by the same representation of the product of scaling groups.[7] Defining a vector field mathematically correctly requires that students think about placing a vector space at every point in space. This requires what's known as describing space as a manifold and creating a tangent bundle. Yikes!

Physicists of course *don't* need that fancy math and we don't teach it even in majors' classes. It's obvious that it makes no sense to equate a distance and a time (unless you assume a fixed speed of light). It's obvious what it means to have a vector at every point in space. It's like a weather map of wind speed. True. But the critical element in being able to do a unit check or define an E field without the fancy math is **building the blend** — mentally combining physical knowledge with mathematical representations. That blending is neither obvious nor easy and needs to be taught.

## Am I asking too much?

As an experienced instructor in introductory physics, you may be feeling some distress at this point. "My curriculum is already packed to the gills! I don't have time to teach a whole bunch of new content!" Agreed. But I am not suggesting adding new content. Rather, I am saying that if we feel that it is important for students to learn to use math in science effectively then we have to teach the content that we teach in a different way.

These new learning goals don't add to the content of the class; they appear in *all* contents, identifying strategies that have general utility. They are *threads*, a way-of-thinking that crosses the content that runs throughout the class, tying it together with technique and modes of thought, like the warp and weft in a weave (Fig. 1).

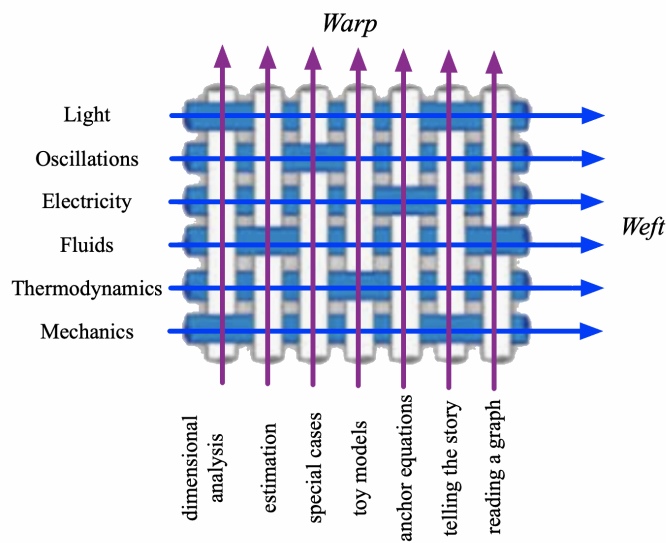

*Figure 1. The interleaving of content (weft) with general strategy learning (warp)*

Later in the paper, I will present a brief list of some of the tools I use to teach mathematical thinking in multiple contexts. But first, we need a deeper understanding of the barriers to students learning these skills.

## Analyzing how students think

When we ask students to do the kind of "thinking with math" that I'm describing, we often not only see them struggle to do it, we often find strong resistance. In a series of interviews with life science students in an introductory physics class, Watkins and Elby[8] documented one student's views of math in science.

> *Ashlyn: I don't like to think of biology in terms of numbers and variables. I feel like that's what physics and calculus are for. So, I mean, come time for the exam, obviously I'm gonna look at those equations and figure them out and memorize them, but I just really don't like them.*

Ashlyn expresses considerable discomfort with trying to use the equations for anything more than calculating.

On the other hand, in another context, she expresses delight in learning in her biology class (with models and math) that the functional dependence of volume on size explains the reason we can't have giant ants that she had previously seen in a Bill Nye episode.

> *Ashlyn: I knew that it doesn't work when you make little things big, but I never had anyone explain to me that there's a mathematical relationship between that, and that was really helpful to just my general understanding of the world. It was, like, mind-boggling…. It was really, like, it blew my mind.*

I've seen both of Ashlyn's reactions from many of my students; resistance to using math to think with and yet delight when the math lets them understand something they knew but didn't understand why it was so.

To understand these complexities of student responses, we need to seek the root causes of students' resistance. I draw on the language of the *resources theoretical framework*, but a detailed knowledge of the framework is not necessary to follow my analysis. I'll give references for those interested in more detail in "Digging deeper" below.

The key idea is that knowledge comes in basic pieces learned from experience in everyday life as well as through schooling. These bits and pieces of knowledge are dynamically activated in the brain in response to both external and internal stimuli.

Some of the differences between math in math and math in science that I've listed above are about expectations about the nature of physics knowledge — *epistemology* — and some are about the kind of things we are talking about — *ontology*.

***Knowledge about knowledge — epistemology:*** In thinking about student difficulties with math in physics, it helps to think about student's *expectations* about the nature of the knowledge they are learning. These are internal stimuli that guide and constrain how students respond to learning in our classes. These expectations may include ideas like, "I know this class is about memorizing equations. I just have to find which equation has the right symbol in it." or "I know I have bad intuition about physics so I have to trust my math even if the result looks crazy." I refer to this kind of expectation as *epistemology* — knowledge about knowledge.

Students' epistemological expectations can have profound effects on what they hear and how they think about what they're learning. I often ask a short essay question on my exams. This helps me get an idea of how the students are thinking beyond "right or wrong". Once, in my early days teaching algebra-based physics, at the end of the first semester (Newtonian mechanics), I asked this on the final exam: "What do you think is the most important equation you learned in this class?" To my embarrassment, the most common answer was $s = 1/2\ at^2$. Newton's 2$^{nd}$ law was rarely mentioned. Clearly I needed to change the way I taught about equations. My students were seeing equations as purely calculational tools, not as ways to help them think about physics or to organize their conceptual knowledge. I totally changed my presentation to focus on Newton's laws as a framework for modeling. I discuss this in detail in the papers in this series on Anchor Equations and Toy Models.

A useful way to think about students' epistemological expectations is to consider the basic ways that they decide they know something: their *epistemological resources*. These can be things they learned as infants ("I know because someone I trusted told me") or that they learned in school ("I know because I solved the equation and got this answer.") Some epistemological resources we want our students to learn in physics include:

- **Symbolic manipulation can be trusted** — Algorithmic transformational steps starting from an application equation lead to a trustable result.

- **Physical meaning can be mapped to math** — A mathematical symbolic representation faithfully characterizes some feature of the physical or geometric system it is intended to represent.

- **If the math is the same, the analogy is good** — Mathematics has a regularity and reliability that is consistent across different applications.

- **Toy models give insight** — Highly simplified examples can yield insight into complex mathematical representation.

- **Consistency is crucial** — When you look at a problem in different ways you should get the same result. This is particularly valuable when a physical and mathematical perspective can be brought independently, confirming an answer in the blend.

Note that epistemological resources, like resources about physical knowledge, can contradict each other. "More means more" can correctly imply "a bigger object has more inertia and is harder to move", but "Closer is stronger" implies that less distance produces a bigger effect such as "closer to a fire is hotter," so in this case, "less is more." The trick to using resources in solving problems is building coherent coordinated collections of them and developing an understanding of the contexts in which each is appropriate.

*Knowledge about things — ontology:* A second issue that is helpful in thinking about student difficulties with math in physics is *ontology* — their knowledge about what kind of things we are talking about.

- Many of our concepts in physics divide something that physically feels like a single concept into two mathematical representations.

For example, students often treat "motion" as a single thing, failing to distinguish velocity from acceleration. While they have good physical experience with both motion and acceleration (catching a ball, feeling a backward push when a vehicle accelerates), building the connection to rates of change is not automatic.

- Some concepts in physics don't match simple everyday ontologies (e.g., matter or process) but require blends.

Nobody promised us that physics was going to be just common sense codified! Quantum physics is the most blatant example, where our description involves characteristics that make objects look both like a particle (localized emission and absorption, momentum conservation in collisions, …) and a wave (interference, uncertainty principle, …). In fact, there are many examples of this, some of them explicit — like treating light as rays, waves, or photons — and some implicit — like considering energy in its matterlike aspects (conserved, has to be positive) and in its representational aspects (can be negative, treat like a location on a graph).[9]

- Some of our concepts in physics are defined mathematically and may not have an obvious physical match.

Particle physicists will be familiar with the fact that fundamental fields (or "elementary particles") are specified as to "what they are" by specifying how they respond mathematically to various transformations — the rotation group (What's its spin?), particle exchange (Is it a fermion or a boson?), and the group of the Standard Model, $U(1) \times SU(2) \times SU(3)$ (What specific particle is it?)

This same sort of "ontology as mathematics" happens in introductory physics as well. We just tend not to notice it. It becomes obvious when we try to explain what an electric field is without discussing vector fields; or when we try to explain why crossed linear polarizers let light through when an angled linear polarizer is inserted between them without talking about vectors and their decomposition. You might be able to do this (I can't), but it feels like trying to play charades with your hands tied behind your back.

# A math-in-science toolbelt

This analysis of the difficulties students have blending math with science raises a lot of questions. Epistemological and ontological issues are *meta* — they run through much of the specific content topics we teach. When we only focus on the list of physics topics we are "covering", we may miss the more general skills that we want students to develop along the way. Physics majors manage to learn these skills, but often over many years (or even decades).

Understanding these deeper issues about the nature of physics knowledge and concepts is even more important to our non-major physics students than any specific content we might choose to teach. But our "I-only-have-to-take-one-year-of-physics" students tend to be focused on "just making it through" and, if they are pre-meds, making sure they get a good score on the exams. Very few will make the effort to think deeply about what they are learning. If we value learning to use math as a scientist, we have to find ways to explicitly teach it to our students.

The key is to teach introductory physics students to make the transition from math as purely about numbers to math as a tool to think about physics. To do this, I've developed a set of general purpose strategies I call *epistemic games* or *e-games* for short: specific approaches that can be brought to bear in a variety of problems. I describe learning to do these as **developing your mathematical toolbelt**. I not only teach these methods explicitly, but each method has a specific tool icon. Every time I use a tool in class, its icon appears on the slide. Every time it's used in our text (a free web-based wiki[10]) the icon appears. As new e-games appear, new problems using them appear on clicker questions, quizzes, and exams.

Here are some of the e-games that my team and I have developed. In each paper in this series, I'll pick one, show how it relates to building an understanding of the use of math in science, and give examples and links to resources.

1. Dimensional analysis
2. Estimation
3. Anchor equations
4. Toy models
5. Functional dependence
6. Reading the physics in a graph
7. Telling the story

Some of the icons used in the NEXUS/Physics materials for these e-games are shown in figure 2.

# Instructional resources

Many of the ideas for this series of paper were developed in the context of studying physics learning in a class for life-science majors. Materials from this project for delivering

9/24/20

instruction on the use of math in physics are available at the *Living Physics Portal*,[11] search "Using math in physics."

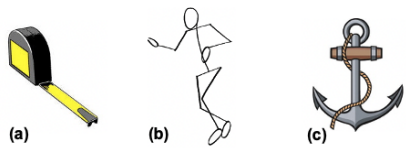

*Figure 2. Some e-game icons from NEXUS/Physics: (a) Dimensional analysis, (b) toy models, and (c) anchor equations.*

## Digging deeper: Research resources

If you're interested in digging into the physics education research literature a bit to see how the issues in this paper are studied, take a look at Redish[12] or Redish & Kuo.[13] (Ref 14 discusses student surprising responses to the equation $E = F/q$ discussed at the beginning of the paper.) The knowledge-in-pieces idea that led to the resources theoretical framework comes from diSessa's phenomenological primitives.[14] The basic principles of the Resources Framework are outlined in Hammer,[15] my Fermi Summer School lecture[16] (available as a preprint on ArXiv), and Hammer et al.[17] The paper introducing the idea of epistemological resources is Hammer & Elby.[18] The idea of epistemic games was introduced by Collins and Ferguson.[19] To see how epistemic games and epistemological resources are observed, see Tuminaro and Redish[20] for a study at the algebra-based level, or Bing and Redish for a study at the advanced physics level. For some papers on the ontological issues, see Gupta, Hammer, and Redish[21] or Dreyfus et al. The Feynman-Hellman theorem example is discussed in detail in Bing and Redish.[22] The general theory of meaning making through blending comes out of the cognitive linguistics literature[23] and is discussed in Dreyfus et al. *loc. cit.*

## Acknowledgements

I would like to thank the members of the UMd PERG over the last two decades for many discussions on these issues, especially Mark Eichenlaub, Ayush Gupta, David Hammer, and Deborah Hemingway. The work has been supported in part by a grant from the Howard Hughes Medical Institute and NSF grants 1504366 and 1624478.

---

[1] E. Redish, Using math in physics - 1. Dimensional analysis, preprint.

[2] This is especially true since word problems were purged from the elementary school mathematics curriculum.

[3] E. Redish, Using math in physics - 3. Anchor equations, preprint.

[4] E. Redish, Using math in physics - 5. Functional dependence and scaling, preprint.

[5] Even my physics graduate students sometimes get queasy when asked to differentiate with respect to a universal constant (like $h$, as in the Feynman-Hellman theorem in a graduate quantum class).

[6] Even introductory math requires a blending of every-day sense with symbol, but in a more abstract and context-independent way. See B. Sherin, How students understand physics equations. *Cognition and Instruction*, **19** (2001) 479–541.

[7] This is really what's going on with unit checks. I'll explain this in the dimensional analysis paper, ref. .

[8] J. Watkins and A. Elby, Context dependence of students' views about the role of equations in understanding biology, *Cell Biology Education - Life Science Education* **12** (June 3, 2013) 274-286. doi:10.1187/cbe.12-11-0185.

[9] B. Dreyfus, A. Gupta, and E. Redish, Applying Conceptual Blending to Model Coordinated Use of Multiple Ontological Metaphors, *Int. J. Sci. Ed.* **37**:5-6 (2015) 812-838.

[10] *The NEXUS/Physics wiki*, https://www.compadre.org/nexusph/

[11] *The Living Physics Portal*, https://www.livingphysicsportal.org/

[12] E. Redish, Analysing the Competency of Mathematical Modelling in Physics. In: T. Greczyło & E. Dębowska (eds) *Key Competences in Physics Teaching and Learning*, Springer Proceedings in Physics, vol 190 (2018).

[13] E. Redish and E. Kuo, Language of physics, language of math: Disciplinary culture and dynamic epistemology, *Science & Education*, **24**:5-6 (2015-03-14) 561-590. doi:10.1007/s11191-015-9749-7

[14] A. diSessa, Towards an epistemology of physics. *Cognition and Instruction*, **10**:2-3 (1993) 105-225;

[15] D. Hammer, Student resources for learning introductory physics, *PER Supplement to the Am. J. Phys.*, **68** (2000) S52-S59;

[16] E. Redish, A Theoretical Framework for Physics Education Research: Modeling student thinking, in *Proceedings of the International School of Physics, "Enrico Fermi" Course CLVI*, Varenna, Italy, August 2003, E.Redish & M. Vicentini (eds.) (IOS Press, 2004).

[17] D. Hammer, A. Elby, R. Scherr, & E. Redish, Resources, framing, and transfer, in J. Mestre (ed.), *Transfer of Learning: Research and Perspectives*, (Information Age Publishing, 2004).

[18] D. Hammer & A. Elby, On the form of a personal epistemology. In B. Hofer & P. Pintrich (eds.), *Personal*